\documentstyle[preprint,prl,aps,epsf]{revtex}
\newcommand{\bm}[1]{\mbox{\boldmath$#1$}}
\newcommand{\lapprox}{\mbox{\raisebox{-4pt}{$\,\buildrel<\over\sim\,$}}}
\begin{document}
\draft
\title{Rotational levels in quantum dots$^*$}
\author{W.~H\"ausler}
\address{I.~Institut f\"ur Theoretische Physik, Universit\"at Hamburg,
D-20355 Hamburg, Germany}
\date{Date: November~2, 1999}
\maketitle
\begin{abstract}
Low energy spectra of isotropic quantum dots are calculated in
the regime of low electron densities where Coulomb interaction
causes strong correlations. The earlier developed pocket state
method is generalized to allow for continuous rotations.
Detailed predictions are made for dots of shallow confinements
and small particle numbers, including the occurance of spin
blockades in transport.
\end{abstract}
\pacs{PACS numbers: 73.20.Mf, 73.50.-h, 73.61.-r}

\narrowtext

\renewcommand{\thefootnote}{\fnsymbol{footnote}}
\footnotetext[1]{Dedicated to Alfred H\"uller to the occasion
of his 60.-{\em th} birthday}

\renewcommand{\baselinestretch}{1.3}\large\normalsize

Much of our present understanding of small quantum dots, with
observable discrete level structure \cite{tarucha,leo},
concentrates on the regime of relatively high carrier densities
where the interaction and charging energy is comparable to
the kinetic (Fermi) energy in magnitude
\cite{merkt,pannkuche,koskinen,suhrke.epl}. Similar to real
atoms effective single particle orbitals establish a reasonable
approximation to the electronic states. The spins follow from
Hund's rule \cite{koskinen,suhrke.epl} which is a perturbative
result though it accords well with experimental findings in
small quantum dots at high particle densities \cite{tarucha}.

At lower densities Coulomb interaction is expected to
destroy this single particle picture, leaving strongly
correlated or even crystallized electrons with collective low
energy excitations. While in the homogeneous two-dimensional
case $\:r_{\rm s}\:$ should exceed $\:r_{\rm c}=37\:$ to reach
this regime \cite{ceperly} ($\:r_{\rm s}=(\pi n_{\rm
s})^{-1/2}\:$ measures the ratio between Coulomb and kinetic
energy and is regulated by the two dimensional carrier density
$\:n_{\rm s}\:$), disorder is predicted to reduce this value
considerably to $\:r_{\rm c}=7.5\:$ \cite{tanatar}. An even
more pronounced reduction of $\:r_{\rm c}\:$ in comparison with
the homogeneous value is found for the transition into the
`Wigner regime' in quantum dots \cite{polyg,mlbdots}. Careful
quantum Monte Carlo (QMC) studies based on the spin sensitivity
of the density--density correlation function yielded $\:r_{\rm
c}=4\:$ for parabolic quantum dots \cite{mlbdots}.
Experimentally, this regime has been addressed using capacitance
spectroscopy \cite{ashoori} which only probes ground state
energies. Non-linear transport behaviour
\cite{tarucha,leo,leo1,weis} has not yet been investigated to
detect the interesting correlation effects for the low energy
excitations.

Numerical investigations of the low density regime, emphasizing
the spin states of rotating three electron Wigner molecules,
have been carried out for shallow parabolic dots \cite{ruan}.
Investigations for larger particle numbers have focussed
on dots of low symmetry where corners in the confining potential
or impurities suppress zero modes to delocalize the charges in
the Wigner regime by so that `pocket states' can be introduced
\cite{fkprobleme}, which are well suited to describe localized
charges. The `pocket states' served as basis to map the spin
sensitive low energy physics to the one of lattice models of the
Hubbard form \cite{jjwh} that account for quantum correlations
by hopping between nearest places. Applicability of this
archetype for correlation phenomena has been demonstrated e.g.\
in quantum dots of polygonal geometry \cite{polyg}.

This mapping to a lattice model cannot be carried out
straightforwardly if zero modes cause charge delocalization
which by symmetry actually happens in most experimental
quantum dots. They are fairly well described by an isotropic
and in fact parabolic model \cite{merkt,stern}
\begin{equation}\label{hdot}
H=\sum_{i=1}^N\frac{\bm{p}_i^2}{2m^*}+V
\end{equation}
where
\begin{equation}\label{pot}
V=\frac{m^*}{2}\omega_0^2\sum_{i=1}^N\bm{x}_i^2+
\sum_{i<j}\frac{e^2}{\kappa|\bm{x}_i-\bm{x}_j|}
\quad.
\end{equation}
Here, the effective mass $m^*$ and the dielectric constant
$\kappa$ are material parameters, and $\bm{x}_j \; (\bm{p}_j)$ are
electron positions (momenta) in two dimensions. This model does
not explicitly involve spin (as opposed to real atoms spin-orbit
coupling is negligible in quantum dots) so that all of its
eigenstates are simultaneously eigenstates to the square of
the total spin $\hat{S}^2$ to eigenvalues $\:S(S+1)\:$. The
present work extends the pocket state method (PSM) to allow for
rotational symmetry and compares with results obtained by QMC
studies \cite{mlbdots}. Being based on a recently developed
multilevel blocking algorithm \cite{mlb} to circumvent the
infamous Fermion sign problem this QMC allows for high accuracy
to resolve reliably even the low energy spin structure at
particle numbers significantly larger than those treatable by
diagonalizations.

At low densities the charge carriers form a finite piece of an
electron crystal \cite{ruan}, a Wigner molecule (WM), that
might, classically \cite{bedanov}, be arbitrarily oriented.
Superposition of all of the azimuthal degeneracies leads to an
isotropic charge density distribution, as required by the
symmetry of (\ref{pot}) \cite{wingreen99}. For analytical
progress it is tempting to separate out the normal coordinate
related with the overall rotation and with total angular
momentum quantum numbers $\:\ell\:$ (in strictly harmonic
confinements $\:\ell\:$ refers to the relative part of the
Hamiltonian since the center of mass motion just adds integer
multiples of $\:\omega_0\:$ to all of the eigenvalues and does
not affect the spin of any of the states \cite{kohn,hawrylak}).
However, the remaining normal coordinates then would in general
no longer describe identical quantum particles obeying Pauli's
principle and Fermi (or Bose) statistics but they would
correspond to linear combinations of such particles. Within the
PSM it is crucial to know the result of particle permutations in
order to assign eventually the correct total spins $\:S\:$ to
the eigenstates and eigenenergies \cite{zpb}.

Therefore, we treat all of the possible particle exchanges on
equal footing, including discrete overall rotations of the WM if
they correspond to particle permutations. It depends on the
geometry of the WM whether rotations by $\:2\pi/p\:$ with
$\:p>1\:$ leave electron places invariant so that the Pauli
principle relates $\:\ell\:$ with $\:S\:$. Such a relationship
is well known, for instance from the example of solid hydrogen
H$_2$, where the even $\ell$ are necessarily $S=0$ singlett
states while the odd $\ell$ are $S=1$ tripletts (in this example
the spins refer to the protons), the reason being the
equivalence of rotations by 180 degrees with the exchange of two
identical spin--half Fermions. Other examples are discussed in
\cite{press}.

Validity of the PSM requires that the spin sensitive excitation
energies $\:\Delta\:$, to be calculated by this method, should
be smaller than charge (plasmon) excitations \cite{zpb}. In the
absence of continuous symmetries this condition is easily
fulfilled at small densities due to the almost exponential decay
of $\:\Delta\sim\exp-\sqrt{r_{\rm s}}\:$. Plasmon energies
decrease only according to a power law $\:\sim r_{\rm
s}^{-3/2}\:$ for Coulomb repulsions. With their faster decay
$\:1/2I=(2\pi m^*\int_0^{\infty}{\rm d}r\;r^3n(r))^{-1}\sim
r_{\rm s}^{-2}\:$ (depending on the radial charge density
distribution $\:n(r)\:$, $I$ is the moment of inertia) the total
angular momentum excitations, however, still decay faster than
the plasmons so that eventually the low energy levels will
follow only from electron interchanges among the places defining
the WM \cite{centrifugal}, including overall rotations by
$\:2\pi/p\:$, i.e.\ by processes permuting identical quantum
particles \cite{annalen}.

From classical \cite{bedanov} as well as from quantum
\cite{mlbdots} Monte Carlo studies it is known that up to
$\:N\le 8\:$ Wigner molecules in the parabolic quantum dots are
very symmetric~: the electrons form one spatial shell ($\:N\le
5\:$) so that $\:p=N\:$, or one electron occupies the center
(i.e.\ $\:p=N-1\:$). Here we focus on $\:N\le 6\:$. The method
can be generalized straightforwardly to larger $N$ and more
complicated geometries of the WM.

The transition amplitudes for all possible particle permutations
constitute the entries $\:t\:$ of the pocket state matrix
\cite{fkprobleme}. In the classically forbidden cases $\:t\:$
can be estimated within the WKB approximation as discussed in
\cite{jjwh,zpb}. The complete potential (\ref{pot}), including
the interaction, goes into this estimate. Often the most
important entries involve only two or three adjacent particles,
as in quantum dots of polygonal shapes \cite{polyg}, which then
determine the hopping terms in the equivalent Hubbard model.
This is different for the zero modes~: there a much larger
number of particles can be involved into a certain permutational
transition, such as a rotation by $\:2\pi/p\:$ in isotropic quantum
dots. Corresponding entries $\:t_{\rm R}\:$ to the pocket state
matrix are not of tunneling type and therefore not exponentially
small. In those cases $\:t_{\rm R}=-p^2/8\pi^2I\:$ is fixed by
the energy constant $\:1/2I\:$ for rotational excitations ($I$
follows from $\:n(r)\:$).

This way all of the relevant entries to the pocket state matrix
can be estimated. Its diagonalization yields eventually the
complete set of low energy eigenvalues. Advantage can be taken
from the fact that pocket states constitute a faithful
representation of the symmetric group $\:S_N\:$ so that
diagonalization can be carried out analytically for small
systems, $\:N\le 4\:$, otherwise numerical help is required.
Only irreducible representations $\:[N/2+S\ ,\ N/2-S]\:$ are
compatible with Pauli's principle for spin-half Fermions
\cite{zpb,haase}. This fixes the spin $\:S\:$ for each eigenvalue.

The entries $|t|\sim{\rm e}^{-\sqrt{r_{\rm s}}}$ and $\:|t_{\rm
R}|\sim r_{\rm s}^{-2}\:$ vary differently with the strength of
the Coulomb interaction so that the ratio $\:t/t_{\rm R}\:$
is a measure for the interaction strength. We use
\[
y:=\frac{1}{1+t/t_{\rm R}}>0
\]
ranging from $\:1/(1+(\pi^2/4)p)\:$, since $\:|2t|\:$ cannot
exceed the Fermi energy in the non-interacting limit, up to
unity at strong interactions, $\:y\to 1\:$.

\begin{figure}\vspace*{12cm}
\includegraphics{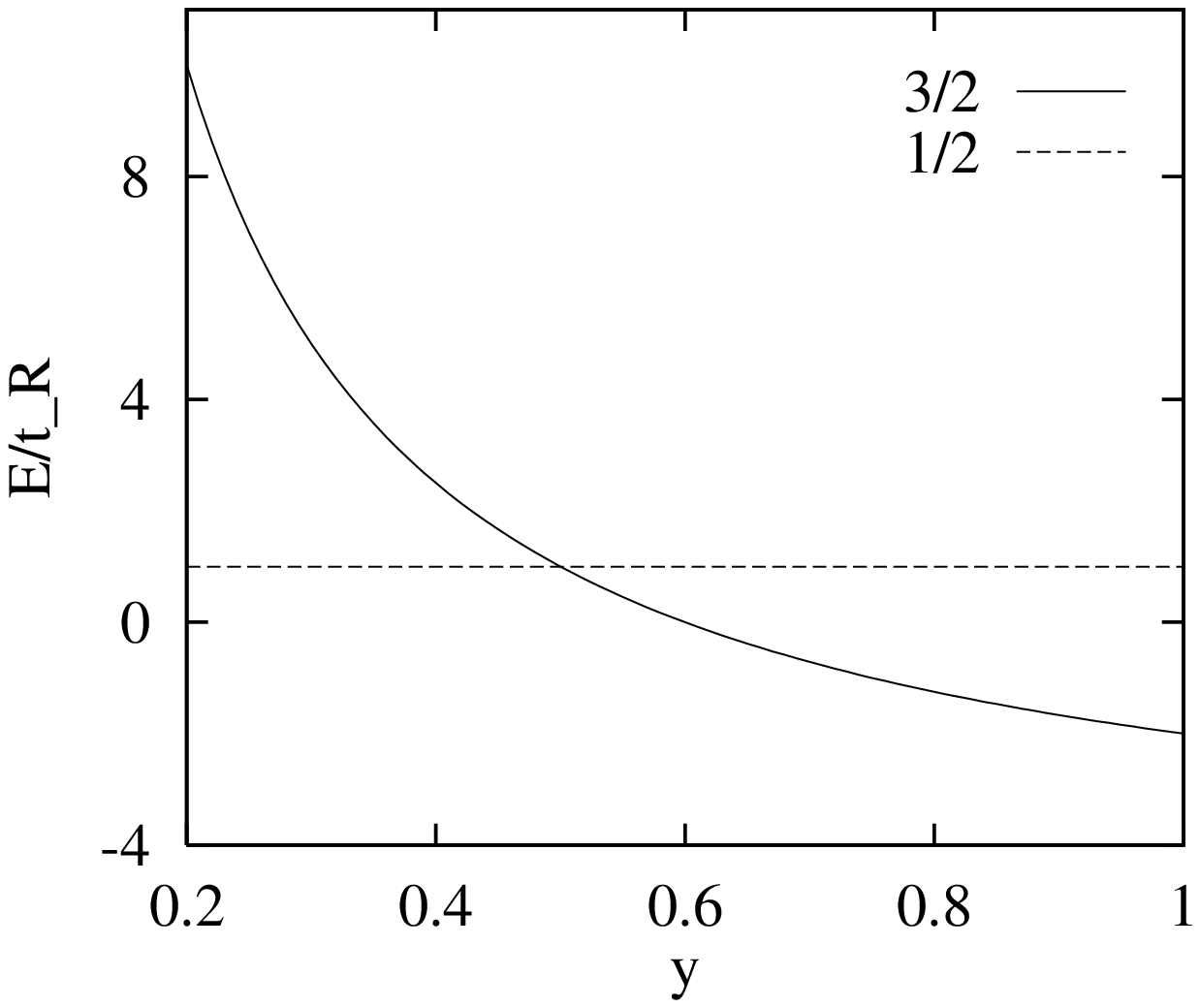}\caption[]{\label{n3}
Low energy levels using pocket states versus $\:y\:$ for
$\:N=3\:$ in units of $t_{\rm R}$.}
\end{figure}

Figure~\ref{n3} shows the low energy spectrum versus $\:y\:$ for
$\:N=3\:$. Our description is designed for evaluating {\em
excitation} energies, i.e.\ the {\em differences} between the
energies of different spin states. As expected for weak
interactions ($\:y<0.5\:$), the ground state is unpolarized
\cite{hawrylak}. A transition into the spin polarized ground
state $\:S=3/2\:$, not found in earlier diagonalization studies,
is seen above a certain interaction strength which for Coulomb
interactions and GaAs parameters can be estimated to happen when
$\:\omega_0<0.5\:$meV \cite{mlbdots}. This result complies with
the QMC studies and can also be seen when carefully examining
Figure~1 of the study \cite{ruan} of a large quantum dot. We
would like to emphasize, that this spin polarization is an exact
consequence of correlations and not the result of the mean field
approximation or a magnetic field. In transport experiments,
when contacting quantum dots with electron reservoirs, it should
show up as a `spin blockade' \cite{wein}, since the ground
states of $\:N=2\:$ and $\:N=3\:$ in sufficiently {\em large}
quantum dots differ then in spin by more than $\:\Delta S=1/2\:$
(by which entering or escaping single electrons can change spin)
since the $\:N=2\:$ ground state (with time reversal symmetry)
is always a singlett \cite{liebmattis}.


For $\:N=4\:$ (not shown here) we confirm the Hund's rule result
of a $\:S=1\:$ ground state, as obtained already in density
functional calculations \cite{koskinen,suhrke.epl,wingreen99}.
New is its persistence up to strong interactions. The lowest
singlett level $\:S=0\:$ approaches this ground level
$\:\sim\exp-\omega_0^{-1/3}\:$ as $\:\omega_0\:$
decreases. The rotationally first excited state $\:\ell=1\:$
consists only of triplett $\:S=1\:$ levels while the spin
polarized level $\:S=2\:$ belongs to the doubly excited
rotational state, $\:\ell=2\:$, together with another singlett
$\:S=0\:$ level.

\begin{figure}\vspace*{12cm}
\includegraphics{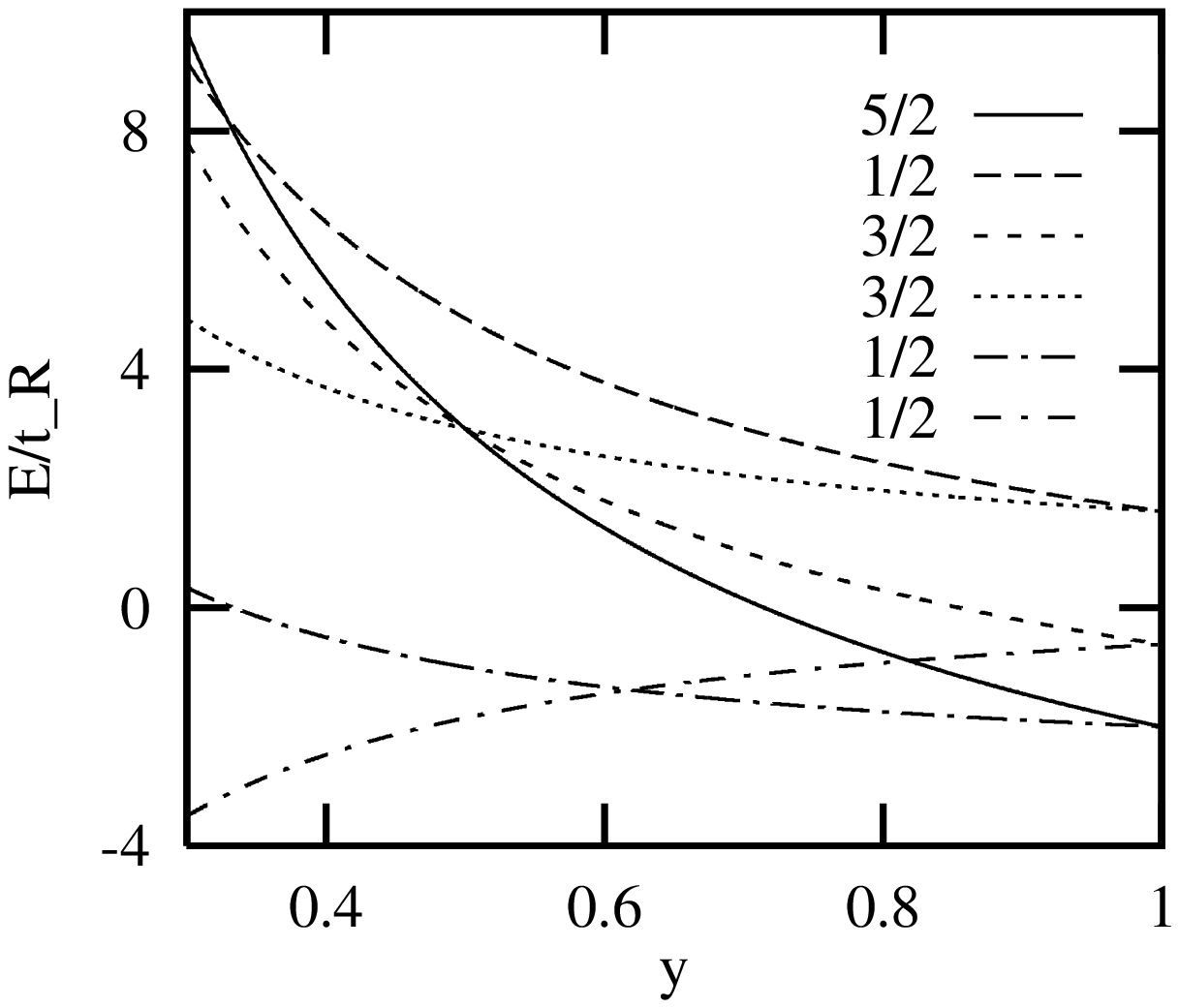}\caption[]{\label{n5}
Low energy levels using pocket states versus $y$ for
$N=5$ in units of $t_{\rm R}$.}
\end{figure}

For $\:N=5\:$ (Figure~\ref{n5}), on the other hand, the
polarized state $\:S=5/2\:$ joins the unpolarized ground state
$\:S=1/2\:$ in the lowest rotational level at strong
interactions. This low energy high spin state makes negative
differential conductances in the non-linear transport likely, due
to the spin blockade \cite{wein}. Rotationally excited levels
consist of $\:S=1/2\:$ as well as of $\:S=3/2\:$ spin states.

The sixth electron is predicted \cite{mlbdots}, also classically
\cite{bedanov}, to occupy the center of a 5--fold ring. This
complicates the pocket state analysis since new types of pair
exchanges appear (exchange with the central electron) and also
the triple exchange $\:t_3\:$ (cyclic permutations of three
adjacent electrons, including the central one) turns out as
important, in accordance with WKB estimates \cite{chakravarty}.
Indeed, the PSM spectra do not compare with the low energy
levels obtained from QMC unless $\:t_3\:$ is included with a
similar magnitude as the pair exchanges.

This demonstrates how our approach complements most favorably
the QMC simulations for quantum dots which yields abolute values
for the many particle energies to high accuracy, contrary to the
method based on pocket states. Very reliable estimates for the
$\:t$--parameters can be achieved which otherwise would have to
be guessed by less trusty approximative means. On the other
hand, QMC is incomplete for the low energy levels since only
the lowest eigenenergies to given $z$--component can be
simulated.

For $\:N=6\:$ and confining energies $\:\omega_0\approx 0.13$meV
(GaAs) we find, with increasing energy, the spin sequence
1-0-3-2-1-0-2-1-2-1-1-0. The spin $\:S=1\:$ indicates another
interaction induced change in the ground state spin since from
the non-interacting levels point of view $\:N=6\:$ corresponds
to a `noble gas' configuration implying an unpolarized ground
state spin $\:S=0\:$ \cite{suhrke.epl}. This result also has to
be contrasted with the conjecture $\:S=2\:$ following from a
static antiferromagnetic WM \cite{jjwh} of pentagonal symmetry.
The rotational ground state $\:\ell=0\:$ includes all possible
spin states $\:S=1,0,3,2\:$, with the fully polarized state,
$\:S=3\:$, being lower in energy than the lowest $\:S=2\:$
state, in accordance with QMC. This again suggests possible
occurance of negative differential conductances for the
transition to $\:N=5\:$.

In conclusion, generalizing the pocket state method we have
developed a description for the low density regime in isotropic
such as parabolic quantum dots. Low energy levels, including
spin quantum numbers were determined for $\:N\le 6\:$. Detailed
predictions are made for spin blockades as they should be
detectable in linear and non-linear transport through shallow
quantum dots of confinement energies below $\:\omega_0\lapprox
0.4$meV (GaAs). \vspace{3mm}.

\noindent{\bf Acknowledgement}\\
I am particularly indebt to my teacher Alfred H\"uller for
longstanding support and encouragement. Numerous very fruitful
discussions with Charles Creffield, Reinhold Egger, Hermann
Grabert, and John Jefferson are acknowledged. This work has been
carried out during stays at the University of Freiburg, the
University of Jyv\"askyl\"a, and the King's College London.
Support has been received from the DFG (through SFB 276) and the
EPSRC (U.K.).

\begin{references}
\bibitem{centrifugal}
For $\:N\ge 3\:$ centrifugal forces are of minor importance; we
disregard them here for simplicity though inclusion would not be
a conceptual problem.
\end{references}
\end{document}